\documentclass[pra,twocolumn,showpacs,letterpaper,showpacs,superscriptaddress]{revtex4}
\usepackage{graphicx,amsmath,amssymb,amsfonts,latexsym,color,dcolumn,bm}

\begin{document}

\newcommand{\beq}{\begin{equation}}
\newcommand{\eeq}{  \end{equation}}
\newcommand{\bea}{\begin{eqnarray}}
\newcommand{\eea}{  \end{eqnarray}}
\newcommand{\bit}{\begin{itemize}}
\newcommand{\eit}{  \end{itemize}}

\title{Sub-Planck phase-space structures and Heisenberg-limited measurements}

\author{F. Toscano}
\affiliation{Instituto de F\'{\i}sica, Universidade Federal do Rio de Janeiro, Caixa Postal 68.528, 21.941-972, Rio de Janeiro, Brazil}

\author{D.A.R. Dalvit}
\affiliation{Theoretical Division, MS B213, Los Alamos National Laboratory,
Los Alamos, NM 87545, USA}

\author{L. Davidovich}
\affiliation{Instituto de F\'{\i}sica, Universidade Federal do Rio de Janeiro, Caixa Postal 68.528, 21.941-972, Rio de Janeiro, Brazil}

\author{W.H. Zurek}
\affiliation{Theoretical Division, MS B213, Los Alamos National Laboratory,
Los Alamos, NM 87545, USA}

\date{\today}

\begin{abstract}

We show how sub-Planck phase-space structures in the Wigner function \cite{Zurek2001}
can be used to achieve Heisenberg-limited sensitivity in weak force measurements.
Nonclassical states of harmonic oscillators, consisting of
superpositions of coherent states, are shown to be useful for the
measurement of weak forces that cause translations or rotations in
phase space, which is done by entangling the quantum oscillator with
a two-level system. Implementations of this strategy in cavity QED
and ion traps are described.

\end{abstract}


\maketitle


\section{Introduction}

Quantum metrology encompasses the estimation of an unknown parameter of a quantum system, and has been the subject of increasing scientific and technological interest due to enhanced measurement techniques allowed by quantum mechanics \cite{Giovannetti2004}.
The two typical problems of small quantum parameter estimation are high precision phase measurements and the detection of weak forces \cite{Gilchrist2004}.
Detection of a small relative phase between two superposed quantum states  includes two equivalent techniques, {\it i.e.}  Ramsey spectroscopy
and Mach-Zehnder interferometry \cite{Yurke1986,Wineland1996}.
They involve detection of a rotation of the quantum state in phase space around the origin.
Thus, the problem of phase determination is ultimately associated with the estimation of a small rotation angle.
Detection of weak forces can be traced back to the pioneering work on gravitational wave detectors that proposed to use a
quantum-mechanical oscillator as an antenna \cite{Braginsky1992,Caves1980}. A weak force (exerted e.g. by the wave)
induces a displacement of the quantum state in phase space in some direction. Thus,
in this case the quantum parameter estimation can be reduced to the determination of a small
linear displacement.

The precision in quantum parameter estimation depends on the energy
resources (e.g., the average number $\bar n$ of photons) involved
in the measurement process. It is well known that using
quasiclassical states the sensitivity is at the standard quantum
limit (SQL), also known as the shot-noise limit. In particular,
coherent states are associated with SQL: The phase space size of a
coherent state is given by $\simeq \sqrt{\hbar}$ and its distance
from the origin is $\simeq \sqrt{\hbar \bar n}$. The smallest
noticeable rotation that will lead to approximate orthogonality is
equal to its angular size as ``seen from the origin", $\sqrt{\hbar}
/ \sqrt{\hbar \bar n} \simeq {\bar n}^{-1/2}$, {\it i.e.}, the
standard quantum limit (see Fig.1). The same argument implies that
the smallest detectable displacement is of the order of
$\sqrt{\hbar}$, so SQL for weak force detection is independent of
$\bar n$, {\it i.e.}, it scales as ${\bar n}^0$. The SQL limit can
be surpassed by using quantum effects (such as entanglement and
squeezing), reaching the so-called Heisenberg limit (HL), in which
the sensitivity is higher than SQL by ${\bar n}^{-1/2}$
\cite{Wineland1996, Holland1993, Milburn2002,Smerzi2005}. Sub
shot-noise sensitivities, approaching the ultimate Heisenberg limit,
can be achieved using path-entangled states of photons
\cite{Angelo2001,Itoh2002,Zeilinger2004,Steinberg2004,Zhao2004,Eisenberg2004,Eisenberg2005}
or ions \cite{Sackett2000,Meyer2001,Liebfried2004}, recently
produced in experiments.

In this paper we show that, as is already anticipated by the brief
discussion of SQL above, the sensitivity of the quantum state to
displacements is related to the smallest phase space structures
associated with its Wigner function $W$. This connection was
conjectured by one of us \cite{Zurek2001} in the context of the
discussion of the sub-Planck structures in $W$. The area of these
structures can be as small as $a=\hbar^2/A$, where $A$ is the action
of the effective support of $W$. $A$ is limited from above by the
classical action of the state, but  it can be much smaller than
that. It is least for a coherent state {\it i.e.} $A \simeq \hbar$, which yields
$a=\hbar^2/\hbar=\hbar$, and then leads to SQL. Sub SQL sensitivities
can be achieved with coherent squeezed states \cite{Caves1981}, that also have
$A \simeq \hbar$ but, contrary to coherent states, have unequal quadratures:
one is contracted $\propto \sqrt{\hbar} e^{- r}$, and the other is
expanded $\propto \sqrt{\hbar} e^{ r}$ ($r>0$ is the squeezing parameter).
Thus, squeezed states have sub shot-noise sensitivity for perturbations acting
along the squeezed direction.  However, for a fixed $\bar n$, we shall show that states with
much larger values  of $A \simeq \hbar \bar n$ can be found, which
exhibits sensitivity set by $\sqrt{a}
\simeq \sqrt{\hbar/\bar n}$ to displacements, which then allows one
to saturate the Heisenberg limit. In this way, we shall demonstrate
that the sub-Planck scale $\hbar^2/A$ determines sensitivity of
small parameter estimation.

Our paper is organized as follows. In the first section we explain
the connection between sub-Planck phase space structures and
Heisenberg-limited sensitivity in quantum metrology. In the second section we discuss
a general scheme for measuring small displacements and rotations in
phase space by using nonclassical states of a harmonic oscillator,
suitably coupled to a two-level system (TLS). We describe in the third section
how to implement our proposal both  in cavity
QED and ion trap experiments, which can take advantage of sub-Planck
structures for quantum-enhanced measurements. Finally, we present
our conclusions.

\begin{figure}[t]
\setlength{\unitlength}{1cm}
\begin{center}
    \scalebox{0.95}[0.95]{%
    \includegraphics*[width=6.2cm]{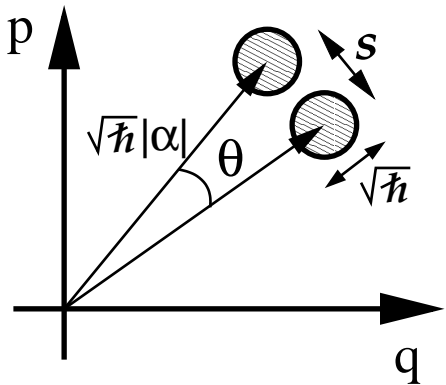}}
 \end{center}
\caption{Phase space representation of the standard quantum limit
(SQL) for rotations. Coherent states have phase space size of the
order $\varepsilon \simeq \sqrt{\hbar}$. A coherent state of complex
amplitude $\sqrt{\hbar}\alpha$ is rotated around the origin by a
small angle $\theta$. The initial and final coherent states are
distinguishable (approximately orthogonal) when the linear
displacement $\varepsilon \approx s \simeq \sqrt{\hbar} |\alpha|
\theta $, which leads to  SQL sensitivity for rotations, $\theta
\simeq 1/ |\alpha|=1/{\bar n}$.} \label{fig1}
\end{figure}


\section{Sub-Planck structures for quantum metrology}

Let us consider  superpositions of $M$ coherent states, equidistantly placed on a circle ${\cal C}$ of radius
$|\alpha| \gg 1$
\beq
\label{state-coh-circle}
|{\rm cat}_M  \rangle= \frac{1}{\sqrt{M}}\sum_{k=1}^{M}
\;e^{i\gamma_k}\;|e^{i\varphi_k}\alpha\rangle\;\;,
\eeq
where $\varphi_k=2\pi k/M$ , and the $\gamma_k$'s are arbitrary phases.
These ``circular states'' are nonclassical states of a harmonic oscillator
for which the mean number of excitations is
$\bar{n}\equiv\langle {\rm cat}_M|\hat{a}^{\dagger}\hat{a}|{\rm cat}_M\rangle
\approx |\alpha|^2$.
States of the form (\ref{state-coh-circle}) include the
periodic case of the ``generalized coherent states'' considered in \cite{Glauber1965,Birula1968}.
These nonclassical states can be generated by nonlinear optical processes \cite{Miranowicz1990,Tara1993}, and by
quantum-nondemolition-measurements of the  photon number in cavity QED \cite{Brune1992} or the
vibrational number of  a trapped ion \cite{Shneider1998}. Some properties of these states were studied in \cite{Janszky1993,Agarwal2005}.
Examples are the Schr\"odinger cat state
\begin{equation}
| {\rm cat}_2 \rangle = (| \alpha \rangle + |-\alpha \rangle)/\sqrt{2} ,
\end{equation}
and the compass state \cite{Zurek2001}
\begin{equation}
| {\rm compass} \rangle =  | {\rm cat}_4 \rangle = (| \alpha \rangle + |- \alpha \rangle +
 | i \alpha \rangle + | -i \alpha \rangle)/2 .
 \end{equation}

We show in the following that when a unitary perturbation $\hat{U}_{x}$ induces a small linear displacement of magnitude $x$, the overlap between the unperturbed state
$|{\rm cat}_M \rangle$ and the perturbed one $| {\rm cat}_M(x)  \rangle= \hat{U}_{x} | {\rm cat}_M \rangle$ oscillates with a typical
frequency $\sim |\alpha|$. Therefore,  the least linear displacement $x=s$ needed to distinguish the two states
is $s \sim 1/|\alpha|$. This scale defines the Heisenberg limit for displacement measurement.
In the case of a rotation, $x=\theta$, quasi-orthogonality occurs when
the rotation induces a linear displacement of the center of the circle ${\cal C}$ of the order of $s \sim \theta |\alpha|$, with $s\sim 1/|\alpha|$.
Therefore, the detectable angle  is $\theta \sim 1/|\alpha|^2$,
defining in this case the Heisenberg limit for
rotation measurements.
We also show that the oscillatory behavior of the overlap function
$|\langle {\rm cat}_M| {\rm cat}_M(x)  \rangle|^2$ has its origin in the
overlap between the oscillatory structure of the Wigner functions of the
unperturbed and the perturbed states whose typical
frequency of their oscillations is precisely $\propto |\alpha|$.
Thus, we prove that the sub-Planck phase space structure of the states in
Eq.(\ref{state-coh-circle}) determines its Heisenberg-limited sensitivity for
quantum metrology applications.

\begin{figure}[t]
\setlength{\unitlength}{1cm}
\begin{center}
    \scalebox{0.95}[0.95]{%
    \includegraphics*[width=6.2cm,angle=-90]{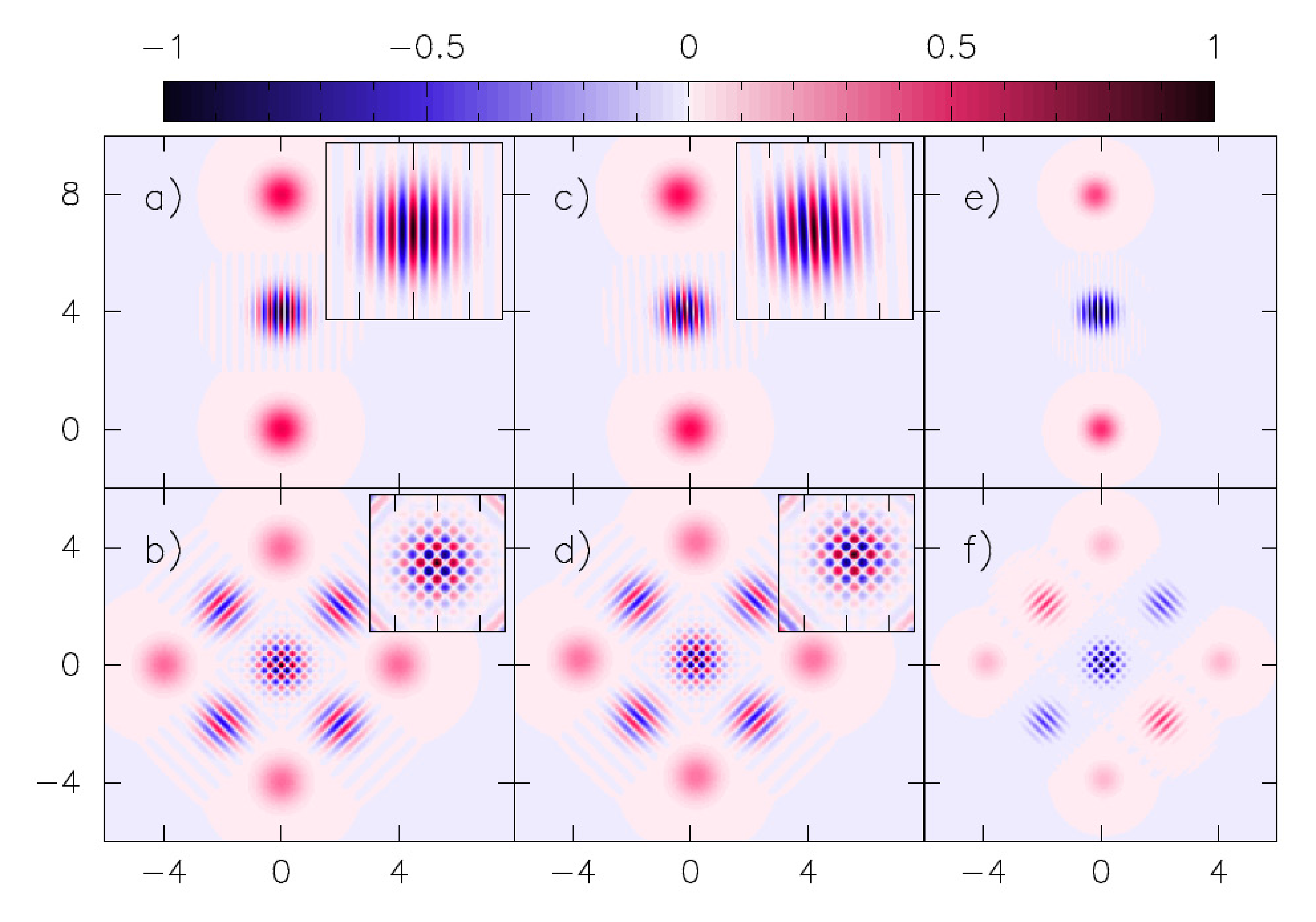}}
\end{center}
\caption{The Wigner functions in the $\alpha$-plane for:
{\bf a)} the displaced ``cat state''
$|\overline{{\rm cat}_2} \rangle\equiv\hat{D}(\alpha)|{\rm cat}_2\rangle$, and
{\bf b)} the ``compass state'' $|{\rm compass}\rangle$ for $\alpha= 4 i$.
The  displaced cat state is quasi-orthogonal
to the rotated state $\hat{R}(\theta)|\overline{{\rm cat}_2} \rangle$ ($\theta=\pi/4|\alpha|^2$) in {\bf c)}
at the Heisenberg limit scale $\theta \sim 1/|\alpha|^2 $.  The compass state is
quasi-orthogonal to the translated compass state $\hat{D}(\beta)| {\rm compass} \rangle$
($\beta=e^{i \pi/4} \pi/2\sqrt{2} |\alpha|$) in {\bf d)}
at the Heisenberg limit scale $|\beta| \sim 1/ |\alpha|$.
The insets enlarge the central interference pattern of the displayed Wigner functions.
In {\bf e)} and {\bf f)} we display the respective products of the unperturbed and perturbed
Wigner functions. When performing the integration over the  $\alpha$-plane,
the negative contributions (in blue) cancel the positive
ones (in red), leading to quasi-orthogonality.
}
\label{fig2}
\end{figure}

{\it Displacements:}
We consider a small linear displacement given by the unitary operator
$\hat{D}(\beta)\equiv e^{\beta\hat{a}^{\dagger}-\beta^*\hat{a}}$ in an  arbitrary direction
$\beta=e^{i\varphi}\alpha s/|\alpha|$ with magnitude $|\beta|=s\ll 1$. This operation transforms
the unperturbed state $|{\rm cat}_M \rangle$ into the perturbed state
$|{\rm cat}_M(s)  \rangle \equiv \hat{D}(\beta) |{\rm cat}_M \rangle$. The overlap between these two
states can be calculated as the integral over phase space $\bar{\alpha}$ of the overlap
between their respective Wigner functions
\begin{equation}
|\langle{\rm cat}_M|{\rm cat}_M(s)\rangle|^2=\int \frac{d^2\bar{\alpha}}{\pi}\,
W_{|{\rm cat}_M \rangle}(\bar{\alpha})\,W_{|{\rm cat}_M(s)\rangle}(\bar{\alpha}) .
\label{overlapgen}
\end{equation}
The Wigner function $W_{|{\rm cat}_M(s)\rangle}(\bar{\alpha})$ of the perturbed state is
\begin{equation}
\label{Wigner-function-perturbed-state}
W_{|{\rm cat}_M(s)\rangle}(\bar{\alpha})=\frac{1}{M} \sum_{k=1}^{M}\sum_{l=1}^{M}
e^{i(\gamma_k-\gamma_l)}e^{is a_{kl}|\alpha|}\; W^{s}_{kl}(\bar{\alpha})\; ,
\end{equation}
where $W^s_{kl}$ are the Weyl$-$Wigner functions
\cite{Wigner1984,Balazs1984} corresponding to the operators $|e^{i
\varphi_k} \alpha + \beta \rangle\langle e^{i \varphi_l}
\alpha+\beta|$, and $a_{kl}\equiv
\sin(\varphi-\varphi_k)-\sin(\varphi-\varphi_l)$. The Wigner
function $W_{|{\rm cat}_M \rangle}(\bar{\alpha})$ of the unperturbed
state is obtained from Eq.(\ref{Wigner-function-perturbed-state}) by
setting $s=0$. The resulting Weyl-Wigner functions $W_{kl} \equiv
W^{s=0}_{kl}$ are
\begin{eqnarray}
W_{kl}(\bar{\alpha})&=&2
\exp\left\{-2\left|\bar{\alpha}-\left(\frac{\alpha_k+\alpha_l}{2}\right)\right|\right\}
\times \nonumber \\
&\times& \exp\left\{i\,2\,{\rm Im}\left(-(e^{-i\varphi_k}-e^{-i\varphi_l})\alpha^*\bar{\alpha}\right)
\right\}
\times\nonumber\\
&\times&\exp\left\{i\,2\,{\rm Im}\left(e^{i(\varphi_l-\varphi_k)}|\alpha|^2\right)\right\}\;.
\end{eqnarray}
Therefore, the Wigner function $W_{|{\rm cat}_M \rangle}$ of the unperturbed state consists of
$M$ Gaussian functions $W_{kk}$ centered at the phase space points
$e^{i\varphi_k}\alpha$, plus interference terms $W_{kl}$ ($l\neq k$) which oscillate
with a typical frequency $\propto |\alpha|$ (see Fig.\ref{fig2}).

Expressing the overlap Eq.(\ref{overlapgen}) in terms of the Weyl-Wigner functions and using that
$|\alpha| \gg 1$ we get
\begin{widetext}
\beq
|\langle{\rm cat}_M|{\rm cat}_M(s)\rangle|^2
\approx
\frac{1}{M^2}\left[\sum_{k=1}^{M}
\int \frac{d \bar{\alpha}^2}{\pi}\; W_{kk}(\bar{\alpha})
\;W^s_{kk}( \bar{\alpha})
+
\sum_{k=1}^{M}\sum_{l>k}^{M}  2\Re e
\left(e^{ia_{kl}|\alpha|s}
\int \frac{d \bar{\alpha}^2}{\pi}\; W_{lk}(\bar{\alpha})
\;W^s_{kl}(\bar{\alpha})\right)
\right]
\;\;.
\label{correlation-function2}
\eeq
\end{widetext}
Here we have neglected contributions
$\int \frac{d\ \bar{\alpha}^2}{\pi}\; W_{lk}(\bar{\alpha})
\; W^s_{k^{'} l^{'}}(\bar{\alpha}) \approx {\cal O}(e^{-|\alpha|^2})$ for $l \neq l^{'}$ and
$k \neq k^{'}$. A further simplification can be achieved using the fact that the perturbation
is small, $|\beta| =s \ll 1$. Indeed, in this case we have
$\hat{D}(\beta) | \alpha \rangle = e^{i {\rm Im}(\beta \alpha^*)} |\alpha+\beta \rangle \approx e^{2 i {\rm Im} (\beta \alpha^*)} |\alpha\rangle$, so that the perturbed and unperturbed Weyl-Wigner functions are
related as
\begin{equation}
W^s_{kl}(\bar{\alpha}) \approx e^{i s a_{kl} |\alpha|} W_{kl}(\bar{\alpha}).
\end{equation}
Therefore, the integral in the first term of Eq.(\ref{correlation-function2}) is equal to 1, and the integral of the second term is equal to $e^{i s a_{kl} |\alpha|}$. Finally, the overlap reads
\beq
\label{correlation-function}
|\langle{\rm cat}_M|{\rm cat}_M(s)\rangle|^2\approx
\frac{1}{M^2} \left[ M+ \sum_{k=1}^M \sum_{l>k}^M 2 \cos( 2 s a_{kl} |\alpha| )
\right] .
\eeq
We see that the oscillations in the function $|\langle{\rm cat}_M|{\rm cat}_M(s)\rangle|^2$ come from the overlap between the interference patterns $W_{kl}$
and $W^s_{kl}$ ($l\neq k$) of the  Wigner functions of the
unperturbed and perturbed states respectively.
The typical frequency of these oscillations is proportional to $|\alpha|$, and implies that
the   states $| {\rm cat}_M \rangle$ are Heisenberg-limited sensitive to displacements
($s \sim 1/|\alpha|$).
Similar oscillations when the initial state is a Fock state were discussed in \cite{Ozorio2004}.

{\it Rotations:} Small rotations in phase space, induced by the operator  $\hat{R}(\theta)=e^{i \theta \hat{a}^{\dagger}\hat{a}}$, with $\theta\ll1$, can be treated in a similar way.
It is first necessary to displace the state $|{\rm cat}_M\rangle$ so that the displaced circle ${\cal C}$ contains the origin of phase space. This can be achieved
by considering the displaced state  $|\overline{{\rm cat}_M} \rangle\equiv \hat{D}(\eta)|{\rm cat}_M \rangle$, with $\eta=\alpha$. If we now rotate this displaced
state in an angle $\theta$ around the origin we obtain
\begin{equation}
\hat{R}(\theta) |\overline{{\rm cat}_M} \rangle  \approx
\frac{e^{2 i \theta |\alpha|^2}}{\sqrt{M}}
\sum_{k=1}^M e^{i (\gamma_k + 2 \theta |\alpha|^2 b_k ) }
|e^{i\varphi_k}\alpha + \eta \rangle ,
\label{rotatedisplace}
\end{equation}
where $b_k\equiv\cos(\varphi_k)-\sin(\varphi_k)$. To obtain this equation we
have used that, in the limit $\theta \ll 1/2 |\alpha|$, we have
$\hat{R}(\theta) |\alpha\rangle = | e^{i \theta} \alpha \rangle \approx | \alpha + i \theta \alpha \rangle$, and
$\hat{D}(\beta) | \alpha \rangle \approx e^{2 i {\rm Im} (\beta \alpha^*)} |\alpha\rangle$.
The state given in Eq.~(\ref{rotatedisplace}) is
the same one obtains by applying a linear displacement $\beta$ to
the state $|\overline{{\rm cat}_M}\rangle$ provided that $\beta$ is
orthogonal to $\eta$ ({\it i.e.},  $\beta = i \eta s / |\eta|$), and
has a magnitude $|\beta| = s = |\alpha| \theta \ll 1$. Then,  the
overlap function between the displaced state $|\overline{{\rm
cat}_M}\rangle$ and the corresponding rotated state $|\overline{{\rm
cat}_M}(\theta)\rangle \equiv \hat{R}(\theta) |\overline{{\rm
cat}_M}\rangle$ is
\begin{eqnarray}
|\langle\overline{{\rm cat}_M} |\overline{{\rm cat}_M} (\theta)\rangle|^2 &\approx&
|\langle\overline{{\rm cat}_M} | \hat{D}(\beta) |  \overline{{\rm cat}_M} \rangle|^2 \nonumber \\
&=&  |\langle{\rm cat}_M | \hat{D}(\beta) | {\rm cat}_M \rangle|^2 .
\end{eqnarray}
The last overlap of this equation is given by
Eq.~(\ref{correlation-function}) with $s=\theta|\alpha|$ (see
Fig.~\ref{fig1}). This shows that the displaced states
$|\overline{{\rm cat}_M}  \rangle$ are HL sensitive to rotations
($\theta \sim 1/ |\alpha|^2$).


\section{Measurement strategy}

Let us consider the simplest case with $M=2$, {\it i.e.}, the cat
state $|{\rm cat}_2 \rangle$. After a small displacement
$\beta=i\alpha s/|\alpha|$, in a direction orthogonal to $\alpha$
(which is the direction of maximum sensitivity), the overlap
function according to Eq.~(\ref{correlation-function}) is \beq
\label{overlap-s} |\langle {\rm cat}_2 | {\rm cat}_2(s) \rangle|^2
\approx \left[ 1+\cos\left(4|\alpha|s\right) \right]/2\;\;. \eeq As
we have seen, this is also the overlap function when we consider a
small rotation, of angle $\theta=s/|\alpha|$, applied to the state
$|\overline{{\rm cat}_2} \rangle\equiv\hat{D}(\alpha)|{\rm
cat}_2\rangle= (1/\sqrt{2})(|2\alpha\rangle + |0\rangle)$. We see
that if we could measure these overlap functions, we could determine
the parameters $s$ or $\theta=s/|\alpha|$ at the Heisenberg limit,
{\it i.e.}, with a sensitivity proportional to $1/|\alpha|$ and
$1/|\alpha|^2$ respectively. It should be noted that the $M>2$
generalized coherent states, such as the compass state ($M=4$), have
sub-Planck structures that lead to HL sensitivity for displacements
in any direction $\beta$. The cat state, however, has minimal (zero)
sensitivity for displacements along the direction $\alpha$.

The measurement of the small perturbations can be realized by entangling the system
with a two level system (TLS). The general method is the following:
we initially prepare the oscillator in a large-amplitude coherent state $| \alpha\rangle$,
and the TLS in one of its two states, say in the upper state $| e \rangle$.
The composite system is then evolved during a certain time $t=T$ under a
unitary evolution $\hat{U}$, which includes the interaction of the oscillator with
the TLS as well as possible additional unitary operations acting only
either on the states of the oscillator or on those of the TLS.
The unitary perturbation $\hat{U}_{x}$ is then applied to the oscillator
(assuming that it does not affect the state of the TLS), and finally the unitary evolution
$\hat{U}$ is undone.
The final entangled state of the composite system is
\beq
| \Psi_f \rangle = \hat{U}^{\dagger}(T) \hat{U}_{x} \hat{U}(T) |e,\alpha \rangle =
\sqrt{P_e} |e, \Psi^e_{S} \rangle +
\sqrt{P_g} |g, \Psi^g_{S} \rangle \;\;,
\eeq
where $P_e$ and $P_g=1-P_e$ are the probabilities of measuring the TLS in levels $e$ and $g$, respectively.
The unitary operator $\hat{U}$ must be
such that the intermediate states $|\Psi\rangle=\hat{U}|e,\alpha\rangle$ and
$|\Phi\rangle=\hat{U}_{x}\hat{U}|e,\alpha\rangle$ verify
$|\langle\Psi|\Phi\rangle|^2\approx|\langle {\rm cat}_M|{\rm cat}_M(x)\rangle|^2$.
Given that $| \langle e,\alpha | \Psi_f \rangle |^2=|\langle\Psi|\Phi\rangle|^2$,
the information about the perturbation parameter $x$, contained in the overlap function
$|\langle {\rm cat}_M|{\rm cat}_M(x)\rangle|^2$,
is then translated into the probabilities, {\it i.e.},
\begin{equation}
P_e =1-P_g= \frac{ | \langle e,\alpha | \Psi_f \rangle |^2}{ | \langle \alpha
| \Psi^e_S \rangle|^2}
\approx\frac{ | \langle {\rm cat}_M | {\rm cat}_M(x) \rangle |^2}
{ | \langle \alpha | \Psi^e_S \rangle|^2}\;\; .
\end{equation}

The method proposed above can also be used to measure the Loschmidt echo, which quantifies
the sensitivity of a quantum system to perturbations \cite{Jalabert2001,Karkuszewski2002,Jacquod2002,Cucchietti2003}.


\section{Cavity QED and ion trap implementations}

The strategy of measuring small perturbations on superpositions of coherent states of quantum harmonic oscillators via two-level systems can be implemented in cavity QED and ion trap experiments.
In the following we describe in detail all the necessary steps for the implementation of the cat
state ($M=2$). Similar strategies can be used for higher ($M>2$) generalized coherent states. For
example, the compass state can be generated in ion traps by means of engineering the ion-laser
interaction in order to realize a nonlinear multiquantum Jaynes-Cummings dynamics. This will be
the subject of a future publication \cite{Dalvit2006}.

Let us consider the interaction between a harmonic oscillator mode
and a two level system as given by the Jaynes-Cummings (JC) model
\cite{Jaynes1963}. In a cavity QED scenario \cite{Brune1992,
Raimond2001}, the harmonic oscillator is a single mode of the
quantized electromagnetic field in the cavity and the TLS is a
Rydberg atom with a two-level electronic transition coupled to the
field through the JC evolution. In an ion-trap scenario
\cite{Wineland1998}, the harmonic oscillator corresponds to the
center-of-mass motion of the trapped ion, and it couples to the TLS
(which corresponds to an internal atomic transition) when the ion is
irradiated by a laser. In the following  we adopt the cavity QED
scenario, adding short remarks on issues that may be specific for
trapped-ion implementations.

The coherent dynamics in the JC model is described by the Hamiltonian
\begin{equation}
\hat{H}_{JC}=\hat{H}_A+\hat{H}_F+\hat{H}_{AF} ,
\end{equation}
where $\hat{H}_A=(\hbar\omega_0/2)\hat{\sigma}_z$ is the atomic TLS
Hamiltonian, with $\hat{\sigma}_z\equiv|e\rangle\langle
e|-|g\rangle\langle g|$, and $\omega_0$ is the transition frequency
between the lower $|g\rangle$ and the upper $|e\rangle$ states. The
harmonic field  mode is described by $\hat{H}_F\equiv
\hbar\omega\hat{a}^{\dagger}\hat{a}$ and the interaction Hamiltonian
is $\hat{H}_{AF}\equiv
(\hbar\Omega_0/2)\left(\hat{\sigma}^{\dagger}\hat{a}+
\hat{\sigma}\hat{a}^{\dagger}\right)$ where
$\hat{\sigma}=|g\rangle\langle e|$ and $\Omega_0$ is the vacuum Rabi
frequency. It is more convenient to use the interaction picture with
respect to the free evolution $\hat{H}_A+\hat{H}_F$, so the JC
dynamics is described by \beq \hat{H}^{I}_{AF}=(\hbar\Omega_0/2)
\left(e^{i\delta t}\hat{\sigma}^{\dagger}\hat{a}+e^{-i\delta t}
\hat{\sigma}\hat{a}^{\dagger}\right)\;\;, \eeq where
$\delta\equiv\omega_0-\omega$ is the detuning. Our method applies
both to the dispersive and the resonant regime.

{\it Dispersive interaction:} We assume first a dispersive
interaction, with $\delta \gg \Omega_0 \sqrt{\bar{n}}$, {\it i.e.},
the frequency of the field $\omega$ is far detuned from the
transition frequency $\omega_0$ of the TLS, and we assume that the
atom has three relevant states $|g\rangle$, $|e\rangle$ and
$|i\rangle$, so that the field in the high-Q cavity couples
dispersively with the states $|g\rangle$ and $|i\rangle$,  while
transitions involving $|e\rangle$ can be neglected. A similar level
scheme was adopted in Ref.~\cite{Brune1992}. We start with the atom
in the state $|g\rangle$ and the field in the cavity in a
large-amplitude coherent state $|\alpha\rangle$. Before the atom
enters into the high-Q cavity it passes through a low-Q cavity and
suffers a resonant $\pi/2$-pulse, so it evolves into
$\hat{U}_{\pi/2}|g\rangle=(|e\rangle+|g\rangle)/\sqrt{2}$. The
interaction time between the atom and the field in the high-$Q$
cavity may be adjusted by atomic velocity selection and Stark
shifting the atomic levels, so that the interaction ceases when
these levels become highly detuned from the cavity
mode~\cite{Raimond2001}. In this way, the interaction time $T$ up to
the middle of the cavity is adjusted so that $\Omega_0^2
T/4\delta=\pi$, where $\delta=\omega_{ie}-\omega$ is the detuning
between the frequency of the field in the cavity, $\omega$, and the
frequency $\omega_{ie}$ of the transition $ g \longleftrightarrow
i$. Therefore, $\hat{U}_{JC}(T)|g,\alpha\rangle=|g,-\alpha\rangle$,
while the state $|e,\alpha\rangle$ remains the same. The state of
the system right before the application of the perturbation is
$|\Psi\rangle=\hat{U}_{JC}(T)\hat{U}_{\pi/2}|g,\alpha\rangle$, and
reads \beq \label{inter-disp-1}
|\Psi\rangle=[|e,\alpha\rangle+|g,-\alpha\rangle]\sqrt{2}\;\; . \eeq

Assume now a displacement perturbation, corresponding to the unitary operation $\hat{U}_s=\hat{D}(\beta)$, with $\beta= i\alpha s/|\alpha|$ and
$|\beta|=s\ll 1$, is applied to this state.
Displacements of the cavity field can be induced by injecting into the cavity coherent fields, produced for instance by a microwave generator,
while in the ion-trap setting they can be generated by forces that displace the
equilibrium position of the ion.
For detecting a small rotation, we first apply the displacement operator  $\hat{D}(\alpha)$,
during a time $\Delta t \ll T$, which leads to the state
\beq
\label{inter-disp-2}
|\Psi\rangle=[|e,2\alpha\rangle+|g,0\rangle]/\sqrt{2}\;\;.
\eeq
A small rotation $\hat{R}(\theta)$ of the cavity field can be implemented by  a
percussive dislocation of one of the mirrors of the cavity,  thus changing the frequency of the mode by a small amount during a small time interval.
Alternatively, one may send through the cavity a fast atom, which interacts dispersively with the field, and follows a trajectory that avoids the interaction with the first atom.
In the ion-trap  context, the same kind of perturbation can be implemented by slightly changing the frequency of the harmonic  trapping potential.
Note that for the state in Eq.~(\ref{inter-disp-1})
the overlap function  $|\langle\Psi|\hat{D}(\beta)|\Psi\rangle|^2$ is equal to
$|\langle{\rm cat}_2|{\rm cat}_2(s)\rangle|^2$ given by Eq.~(\ref{overlap-s}).
In an analogous way, for the state in Eq.~(\ref{inter-disp-2}) we have
$|\langle\Psi|\hat{R}(\theta)|\Psi\rangle|^2=|\langle{\rm cat}_2|{\rm cat}_2(\theta)\rangle|^2$,
also given by Eq.~(\ref{overlap-s}) with $s=\theta|\alpha|$.

After the perturbation is applied,
we undo the total unitary evolution $\hat{U}_{JC}(T)\hat{U}_{\pi/2}$ (or
$\hat{D}(\alpha)\hat{U}_{JC}(T)\hat{U}_{\pi/2}$ for a rotation perturbation), by letting the atom interact with the cavity field again for a time $T$. Since $T$ is half the period of the dispersive JC evolution, when
the atom leaves the cavity at time $2T$ the JC dynamics is automatically undone.
Up to a global phase, the final state is
\begin{equation}
| \Psi_f \rangle =\frac{1}{2}  \left( 1-e^{i \,4 \,|\alpha|\, s} \right)  |e, \alpha \rangle +
\frac{1}{2} \; \left(e^{i\, 4\,|\alpha|\,s} + 1 \right) |g, \alpha \rangle \;.
\end{equation}
For a small rotation $\theta$, we obtain the same final state with the
displacement $s$ replaced by $\theta |\alpha|$.

The probabilities that the atom exits the cavity
in the upper and lower state depend on the small parameter $s$
(equivalently $\theta=s/|\alpha|$),
\begin{equation}
\label{dif-proba}
P_e=1 - P_g =[1-\cos(4\,|\alpha|\,s)]/2 \;,
\end{equation}
thus exhibiting the characteristic oscillation associated with the interference pattern of the Wigner function.
A good estimate of the unknown parameter $s$ requires repeating the measurement
several times. After $R$ repetitions, the probability that the outcome $|e \rangle$
is obtained $r$ times is given by a binomial distribution. In the large $R$ limit, it is well approximated
by a Gaussian distribution in the variable $\xi = r/R$, which can be regarded as effectively continuous \cite{Luis2004}. In this limit the probability distribution for the estimator
$\tilde{s}=\arccos(2 r /R -1)/4 |\alpha|$ of the true displacement $s$ is \cite{caveats}
\begin{equation}
P(\tilde{s}) \approx  \frac{1}{\sqrt{2 \pi \Delta \tilde{s}^2}} \; e^{ - \frac{(\tilde{s} -s )^2}{2 \Delta \tilde{s}^2}} ,
\end{equation}
where the uncertainty of $\tilde{s}$ is $\Delta \tilde{s}=1/8 \sqrt{R \bar{n}}$, reaching the Heisenberg precision for displacement since
$R \bar{n}$ is the total number of photons used in the measurement.

{\it Resonant interaction:}
We discuss now the case of resonant coupling, $\delta=0$. This case has over the dispersive  case the advantage of requiring much shorter transit
times. The corresponding experimental setup leads to collapses and revivals of the atomic population \cite{Eberly1980}. We start with an initial  product state of the
TLS-oscillator composite system, $|e, \alpha\rangle$, in which the field coherent state has  a mesoscopic mean number of photons $\bar{n}=|\alpha|^2$.
The joint evolution of the atom-field system inside the cavity is given by $\hat{U}_{JC}\equiv \exp\{-i\hat{H}^{I}_{AF}t/\hbar\}$, and it can be calculated
following the approach developed in \cite{Banacloche1991}. Since the field in the cavity is a superposition of different number of photon states, the corresponding
Rabi frequencies are spread. Therefore, the atom gets entangled with the field in a quantum superposition of two coherent components that rotate in opposite directions
in phase space \cite{Haroche2003}.  We set up the velocity of the atom so that the transit time $T$ up
to the middle of the cavity is half the revival time $T_R= 4\pi \sqrt{\bar{n}}/\Omega_0$. This transit time is much shorter than the one for the dispersive case.  The evolved state $|\Psi\rangle =\hat{U}_{JC}(T)\,|e,\alpha\rangle$ turns out  to be the product state \cite{Banacloche1991,Haroche2003},
\beq
\label{intermediate-state-resonant-displacement}
|\Psi\rangle =
\left[
e^{- i \frac{\pi}{2} \bar{n}} | -i \alpha \rangle - e^{i \frac{\pi}{2} \bar{n}} |i \alpha \rangle
\right]/\sqrt{2} \otimes |\phi\rangle_A
\;\;,
\eeq
where $|\phi\rangle_A\equiv(1/\sqrt{2}) \left(
e^{-i \frac{\pi}{2} } |e \rangle + e^{-i {\rm arg}(\alpha)} |g \rangle
\right)$.

A small displacement is then applied to the field. At this point  the JC dynamics  must be inverted.
This can be done by a procedure developed in \cite{Morigi2002}: one applies a percussive controlled phase kick corresponding
to the unitary operation  $\hat{U}_{\rm kick} = \hat{\sigma}_z$ that changes the sign of the
relative phase between the atomic levels.
This amounts to changing the sign of the interaction Hamiltonian ($\hat\sigma\rightarrow-\hat\sigma$), so the phase kick mimics the time-reversal  operation.
This idea was experimentally implemented  in cavity QED  \cite{Haroche2005} and
can be similarly applied in the context of ion traps.
The final state
$| \Psi_f \rangle=\hat{U}_{JC}^{\dagger}(T)\hat{D}(\beta)\hat{U}_{JC}(T)|e,\alpha\rangle$,
up to a global phase, is
\beq
\label{final}
| \Psi_f \rangle = \frac{1}{2}  \left( e^{i \,4 \,|\alpha|\, s} + 1 \right)  |e, \alpha \rangle +
\frac{b}{2} \, \left( 1- e^{i\, 4\,|\alpha|\,s} \right) |g, \alpha \rangle,
\eeq
where $b\equiv e^{-i {\rm arg}(\alpha)} $.
For small rotations, one proceeds as in the previous case, first displacing the field state in Eq.~~(\ref{intermediate-state-resonant-displacement}), then applying the rotation and subsequently inverting the displacement and the time evolution. With the replacement $s\rightarrow \theta|\alpha|$,
one gets the same final state (\ref{final}). Given this final state, one can easily evaluate the probabilities $P_e$ and $P_g$ that the atom exits the cavity in
the upper and lower level, and conclude that also in the case of resonant Jaynes-Cummings interaction one can measure
weak forces at the Heisenberg limit.

>From an experimental point of view, the resonant case is more
convenient than the dispersive one because the interaction times are
much shorter. One should also note that, instead of applying the
percussive time-inversion pulse, the same result would be obtained
by letting the first atom go away of the cavity, after
disentanglement, and then sending a second atom, prepared in the
``time inverted'' state, obtained from $|\phi\rangle_A$ by changing
the sign of the relative phase between the states $|e\rangle$ and
$|g\rangle$. Further shortening of the interaction time can be
achieved by letting the atom interact with the field for a time
$\Delta t < T_R/2$, so that in the intermediate state the atom is
entangled with the two coherent states $|\alpha e^{\pm
i\phi/2}\rangle$ ($\phi=\Omega_0 \Delta t / 2 \sqrt{\overline{n}}$),
and then inverting the dynamics. After an equal amount of time, one
gets again a state like the one in Eq.~(\ref{final}), with $s$
replaced by $s\sin(\phi/2)$, which implies reduced sensitivity, but
does not change the scaling of the minimum detectable displacements
and rotations.

Finally, we discuss the viability of experimental demonstration with
cavity QED and ion trap implementations. For cavity QED, one should
have the interaction time $T=2\pi\sqrt{\bar n}/\Omega_0$ much smaller than the
decoherence time, given for the low temperatures used in typical
experiments by $\tau_{\rm cav}/{\bar n}$, where $\tau_{\rm cav}$ is
the damping time of the cavity field (this condition is probably too
strict, in view of the fact that the maximum distance in phase space
between the two coherent components of the cat state, and therefore
the maximum decoherence rate, is achieved only when the atom is in
the middle of its trajectory). According to this criterium, one
should have therefore $\tau_{\rm cav}\gg 2\pi(\bar
n)^{3/2}/\Omega_0$. For a typical value $\Omega_0=3\times 10^5$
s$^{-1}$ and $\bar n=20$, one gets that $\tau_{\rm cav}\gg 1.9$ ms.
This condition is within reach of present techniques in cavity QED,
where damping times of the order of 15 ms have already been
achieved~\cite{Haroche}.  Atomic state detection has also been
perfected. Present efficiency is between 80\% and 100\%
\cite{Maioli2005}, which should be sufficient to detect the
sub-Planck oscillations.

For ions, detection efficiency is close to 100\%, but one still has
to consider decoherence effects affecting the vibrational cat state.
Considering a typical value $2\pi/\Omega_0=140 \mu$s, one gets
$T\approx0.6$ ms for a vibrational state with $\bar n=20$. Assuming
a damping time for the center-of-mass motion of 100-200 ms, which is
compatible with present experiments~\cite{Blatt}, the decoherence
time for the vibrational cat state would be between 5 and 10 ms,
thus satisfying the requirement that it should be much larger than
the interaction time $T$.


\section{Conclusions}

We have shown that sub-Planck quantum phase space structures \cite{Zurek2001} have
remarkable implications for quantum parameter estimation, as they are responsible for
Heisenberg-limited sensitivity to perturbations. We have proposed a general method to measure
perturbations with such high sensitivity, coupling a harmonic oscillator with a two level
system. This method was applied to cavity QED and ion-trap settings, which should
be within experimental reach.


\section{Acknowledgements}

We are grateful to R.L. de Matos Filho, J.M. Raimond, and A. Smerzi for fruitful discussions.
We thank partial support from
NSA. FT and LD acknowledge the support of the program Millennium
Institute for Quantum Information and of the Brazilian agencies
FAPERJ and CNPq. FT thanks Los Alamos National Laboratory for the
hospitality during his stay.



\begin{thebibliography}{99}

\bibitem{Zurek2001}
W.H. Zurek,
Nature (London) {\bf 412}, 712 (2001).

\bibitem{Giovannetti2004}
For a recent review, see V. Giovannetti, S. Lloyd, and L. Maccone,
Science {\bf 306}, 1330 (2004).

\bibitem{Gilchrist2004}
A. Gilchrist, K. Nemoto, W.J. Munro, T.C. Ralph, S. Glancy, S.L. Braunstein, and G.J. Milburn,
J. Opt. B: Quantum Semiclass. Opt, {\bf 6}, S828 (2004).

\bibitem{Yurke1986}
B. Yurke, S.L. McCall and J.R. Klauder,
Phys. Rev. A, {\bf 33}, 4033, (1986).

\bibitem{Wineland1996}
J.J. Bollinger, W.M. Itano, D.J. Wineland, and D.J. Heinzen,
Phys. Rev. A, {\bf 54}, R4649, (1996).

\bibitem{Braginsky1992}
V.B. Braginsky and F.Ya. Khalili,
{\it Quantum Measurements} (Cambridge University Press, Cambridge, 1992).

\bibitem{Caves1980}
C.M. Caves, K.S. Thorne, R.W.P. Drever, V.D. Sandberg, and M. Zimmermann,
Rev. Mod. Phys. {\bf 52}, 341 (1980).

\bibitem{Holland1993}
M.J. Holland and K. Burnett,
Phys. Rev. Lett. {\bf 71}, 1355 (1993).

\bibitem{Milburn2002}
W.J. Munro, K. Nemoto,  G.J. Milburn, and S.L. Braunstein,
Phys. Rev. A {\bf 66}, 023819 (2002).

\bibitem{Smerzi2005}
L. Pezz{\'e} and A. Smerzi,
quant-ph/0508158.

\bibitem{Angelo2001}
M. D'Angelo, M.V. Chekhova, and Y. Shih,
Phys. Rev. Lett. {\bf 87}, 013602 (2001).

\bibitem{Itoh2002}
K. Edamatsu, R. Shimizu, and T. Itoh,
Phys. Rev. Lett. {\bf 89}, 213601 (2002).

\bibitem{Zeilinger2004}
P. Walther, J.W. Pan, M. Aspelmeyer, R. Ursin, S. Gasparoni, and A. Zeilinger,
Nature {\bf 429}, 158 (2004).

\bibitem{Steinberg2004}
M.W. Mitchell, J.S. Lundeen, and A.M. Steinberg,
Nature {\bf 429}, 161 (2004).

\bibitem{Zhao2004}
Z. Zhao, Y.-A. Chen, A.-N. Zhang, T. Yang, H.J. Briegel, and J.-W. Pan,
Nature {\bf 430}, 54 (2004).

\bibitem{Eisenberg2004} H.S. Eisenberg, G. Khoury, G.A. Durkin, C. Simon, and D. Bouwmeester,
Phys. Rev. Lett. {\bf 93}, 193901 (2004).

\bibitem{Eisenberg2005} H.S. Eisenberg, J.F. Hodelin, G. Khoury, and D. Bouwmeester,
Phys. Rev. Lett. {\bf 94}, 090502 (2005).

\bibitem{Sackett2000}
S.A. Sackett, D. Kielpinski, B.E. King, C. Langer, V. Mayer, C.J. Myatt, M. Rowe, Q.A. Turchette, W.M. Itano, and D.J. Wineland,
Nature {\bf 404}, 256 (2000).

\bibitem{Meyer2001}
V. Meyer, M.A. Rowe, D. Kielpinski, C.A. Sackett, W.M. Itano, C. Monroe, and D.J. Wineland,
Phys. Rev. Lett. {\bf 86}, 5870 (2001).

\bibitem{Liebfried2004}
D. Liebfried,  M.D. Barrett, T. Schaetz, J. Britton, J. Chiaverini, W.M. Itano, J.D. Jost, C. Langer, and D.J. Wineland,
Science {\bf 304}, 1476 (2004).

\bibitem{Caves1981}
C. M. Caves, Phys. Rev.D {\bf 23}, 1693 (1981).

\bibitem{Glauber1965}
U. M. Titulaer and R. J. Glauber,
Phys. Rev. {\bf 145}, 1041 (1966).

\bibitem{Birula1968}
Z. Bialynicka-Birula,
Phys. Rev. {\bf 173}, 1207 (1968).

\bibitem{Miranowicz1990}
A. Miranowicz, R. Tanas, and S. Kielich,
Quantum Opt. {\bf 2}, 253 (1990).

\bibitem{Tara1993}
K. Tara, G.S. Agarwal, and S. Chaturvedi,
Phys. Rev. A {\bf 47}, 5024 (1993).

\bibitem{Brune1992}
M. Brune, S. Haroche, J.M. Raimond, L. Davidovich, and N. Zagury,
Phys. Rev. A {\bf 45}, 5193 (1992).

\bibitem{Shneider1998}
S. Schneider, H.M. Wiseman, W.J. Munro, and G.J. Milburn,
Fortschr. Phys. {\bf 46}, 391 (1998).

\bibitem{Janszky1993}
J. Janszky,  P. Domokos, and P. Adam,
Phys. Rev. A {\bf 48}, 2213 (1993).

\bibitem{Agarwal2005}
P.K. Pathak and G.S. Agarwal,
Phys. Rev. A {\bf 71}, 043823 (2005).

\bibitem{Wigner1984}
M. Hillery, R. F. O`Connell, M. O. Scully and E. P. Wigner,
Phys. Rep. {\bf 106}, No. 3, 121-167 (1984)

\bibitem{Balazs1984}
N. L. Balazs and B. K. Jennings, Phys. Rep. {\bf 104}, No. 6, 347-391 (1984).

\bibitem{Ozorio2004}
A. M. Ozorio de Almeida,  R.O. Vallejos, and M. Saraceno,
J. Phys. A {\bf 38}, 1473 (2005).

\bibitem{Jalabert2001}
R.A. Jalabert and H.M. Pastawski,
Phys. Rev. Lett. {\bf 86}, 2490 (2001).

\bibitem{Karkuszewski2002}
Z.P. Karkuszewski, C. Jarzynski, and W.H. Zurek,
Phys. Rev. Lett. {\bf 89}, 170405 (2002).

\bibitem{Jacquod2002}
P. Jacquod, I. Adagideli, and C.W.J. Beenakker,
Phys. Rev. Lett. {\bf 89}, 154103 (2002)

\bibitem{Cucchietti2003}
F.M. Cucchietti, D.A.R. Dalvit, J.P. Paz, and W.H. Zurek,
Phys. Rev. Lett. {\bf 91}, 210403 (2003).

\bibitem{Dalvit2006} D.A.R. Dalvit {\it et al.},  in preparation.

\bibitem{Jaynes1963}
E.T. Jaynes and F.W. Cummings,
Proc. IEEE {\bf 51}, 89 (1963).

\bibitem{Raimond2001}
J. M. Raimond, M. Brune, and S. Haroche, Rev. Mod. Phys. \bf 73\rm,
565 (2001).

\bibitem{Wineland1998}
D.J. Wineland, C. Monroe, W.M. Itano, D. Liebfried, B.E. King, and D.M. Meekhof,
J. Res. Natl. Inst. Stand. Technol. {\bf 103}, 259 (1998).

\bibitem{Luis2004}
A. Luis,
Phys. Rev. A {\bf 69}, 044101 (2004).

\bibitem{caveats}
The only prior information about the signal is that $0 \le s \le \pi/4 |\alpha| \ll 1$ for displacements, or
equivalently $0 \le \theta \le \pi/4 |\alpha|^2 \ll 1$ for rotations. This is not a restrictive condition
since one can set up the value of $|\alpha|$ in the experiment in order for $\pi/4 |\alpha|$ to be an upper
bound of the expected displacement (or $\pi/4 |\alpha|^2$ an upper bound for expected rotations) to be measured.

\bibitem{Eberly1980}
J. H. Eberly,  N.B. Narozhny, and J.J. Sanchez Mondragon,
Phys. Rev. Lett. {\bf 44}, 1323 (1980).

\bibitem{Banacloche1991}
J. Gea-Banacloche,
Phys. Rev. A {\bf 44}, 5913 (1991).

\bibitem{Haroche2003}
A. Auffeves, P. Maioli, T. Meunier, S. Gleyzes, G. Nogues, M. Brune, J.M. Raimond, and S. Haroche,
Phys. Rev. Lett. {\bf 91}, 230405 (2003).

\bibitem{Morigi2002}
G. Morigi, E. Solano, B.-G. Englert, and H. Walther,
Phys. Rev. A {\bf 65}, 040102(R) (2002).

\bibitem{Haroche2005}
T. Meunier,  S. Gleyzes, P. Maiolli, A. Auffeves, G. Nogues, M. Brune, J.M. Raimond, and S. Haroche,
Phys. Rev. Lett. {\bf 94}, 010401 (2005).

\bibitem{Haroche} S. Haroche, private communication.

\bibitem{Maioli2005} P. Maioli, T. Meunier, S. Gleyzes, A. Auffeves, G. Nogues, M. Brune, J. M. Raimond, and S.
Haroche, Phys. Rev. Lett. \bf 94\rm, 113601 (2005).

\bibitem{Blatt} R. Blatt, private communication.

\end{thebibliography}
\end{document}